\documentclass[conference]{IEEEtran}

\usepackage[utf8]{inputenc}
\usepackage[T1]{fontenc}
\usepackage{soulutf8}
\usepackage{url}
\usepackage{color}
\usepackage{xspace}
\usepackage{booktabs}
\usepackage{makecell}
\usepackage{multirow}
\usepackage{rotating}
\usepackage{tikz}
\usepackage{calc}
\usepackage{subcaption}
\usepackage{algpseudocode}
\usepackage{amsmath}
\usepackage{amssymb}
\usepackage{mathtools}
\usepackage{mathrsfs}

\title{Exploring the consequences of cyber attacks on Powertrain Cyber Physical Systems}

\author{%
  \IEEEauthorblockN{%
    Dario Stabili\IEEEauthorrefmark{1},
    Raffaele Romagnoli\IEEEauthorrefmark{2},
    Mirco Marchetti\IEEEauthorrefmark{1},
    Bruno Sinopoli\IEEEauthorrefmark{3} and
    Michele Colajanni\IEEEauthorrefmark{4}}
    \\
    
    \IEEEauthorblockA{\IEEEauthorrefmark{1} University of Modena and Reggio Emilia\\
        \IEEEauthorrefmark{2} Carnegie Mellon University\\
        \IEEEauthorrefmark{3} Washington University in St. Louis\\
        \IEEEauthorrefmark{4} University of Bologna\\
        \\
        Emails: \IEEEauthorrefmark{1}\{dario.stabili, mirco.marchetti\}@unimore.it,
                \IEEEauthorrefmark{2}rromagno@andrew.cmu.edu\\
                \IEEEauthorrefmark{3}bsinopoli@wustl.edu,
                \IEEEauthorrefmark{4}michele.colajanni@unibo.it}
}

\begin{document}
\maketitle
\thispagestyle{empty}
\pagestyle{empty}

\begin{abstract}
This paper proposes a novel approach for the study of cyber-attacks against the powertrain of a generic vehicle. The proposed model is composed by a a generic Internal Combustion engine and a speed controller, that communicate through a Controller Area Network (CAN) bus. We consider a threat model composed by three representative attack scenarios designed to modify the output of the model, thus affecting the rotational speed of the engine. Two attack scenarios target both vehicle sensor systems and CAN communication, while one attack scenario only requires injection of CAN messages. To the best of our knowledge, this is the first attempt of modeling the consequences of realistic cyber attacks against a modern vehicle.
\end{abstract}
\section{Introduction}
\label{s:introduction}
Since the introduction of microcontrollers in modern vehicles, car manufacturers began to push more and more features to improve both safety and driving comfort. These features are deployed on microcontrollers that are part of the vehicle network and called Electronic Control Units (ECUs), which are connected to the mechanical parts of the vehicle and communicate between each others through means of different communication protocols. The most deployed communication protocol is the Controller Area Network (CAN), developed by Bosch GmbH in the early `90~\cite{BOSCHCanSpecV2}. 
Despite the CAN bus is one of the most deployed networking protocols in modern vehicles it does not provide any security guarantees. Security researchers demonstrated that it is possible to hijack a vehicle by injecting maliciously forged messages on the CAN bus, and published these results on technical reports and white papers~\cite{MillerValasekAdventures, KeenSecurity}. 
These attacks are achieved by exploiting the \textit{drive-by-wire} capabilities of modern vehicles, a feature that enables the control of the driving system through messages sent over the CAN bus. An example of the \emph{drive-by-wire} capabilities of modern vehicle is represented by the cruise control, which is activated to preserve the speed of the vehicle to reduce fuel consumption and therefore emissions.

Despite these systems are developed for comfort and safety purposes, the fundamentals required for their deployment pave the way to targeted attacks, jeopardizing the safety of people inside and outside the vehicle. 
Since the first public demonstration of a remote attack to a modern connected vehicle~\cite{MillerValasekBlackHatPaper}, many cyber security researchers proposed different detection algorithms tailored for the in-vehicle communication networks to detect an ongoing cyber-attack~\cite{Marchetti2017Sequences, Kneib2018Scission, Stabili2021ITASEC}.
However, current literature explored attacks and their relative countermeasures only focusing on a particular vehicle domain. As an example, the detection algorithms designed to detect cyber-attacks targeting the communication network of modern vehicles are tested only against attacks targeting these networks, while detection methods based on the analysis of the system state are only tested against attacks designed to modify the system state.

In this paper we present a complete powertrain system of a generic vehicle, composed by an internal combustion engine and speed controller, connected using a Controller Area Network (CAN) communication protocol. This system is used to analyze the consequences of cyber-attacks to the powertrain system targeting both the sensors and the communication network.

This paper has two main contributions. The first one is that the speed controller is designed and implemented by using the model of the engine instead of relying on a generic representation of the whole vehicle~\cite{Tianxiang2017DoSCPS}. To the best of our knowledge, this is the first time that the speed control problem is addressed by proposing a solution based on the engine model. 
The second contribution of this work is the demonstration of the consequences of cyber-attacks to the engine model, considering different attack scenarios that can be deployed at both system and communication level. To the best of our knowledge, this is the first paper that addresses the cyber security of the engine model using both model and communication methodologies.

The rest of the paper is organized as follows. Section~\ref{s:related} presents an analysis of the current state-of-the-art with the methodology presented in this paper, while Section~\ref{s:powertrain} shows the design of the powertrain system, composed by the engine and controller models and the CAN messages used for communication between the two models. Section~\ref{s:threat_model} introduces the threat model considered in this paper, and its effects on the engine model are presented in Section~\ref{s:attack_consequences}. Final remarks and future work are outlined in Section~\ref{s:conclusions}.

\section{Related Work}
\label{s:related}
Modern vehicle systems are built by considering many different complex domains, that can be grouped as either mechanical~\cite{guzzella2009introduction, Kirli2019Torque}, electronic~\cite{Patel2009Electronics}, or intercommunication domains~\cite{BOSCHCanSpecV2, Ruff2003LIN}.

With the advent of remote connectivity, these systems transitioned from being isolated from outside communication to being interconnected with different communication technologies. The increasing adoption of novel Cooperative Intelligent Transport Systems (C-ITS) paved the way to the development of novel communication standards designed to meet the strict requirements of modern vehicles, targeting both safety and security of the communication~\cite{SecOCSpecification, Bella2020Cinnamon}.

However, these novel external communication methods exposed internal and isolated networks to remote attacks, which take advantage of insecure networks to hijack the vehicle system~\cite{MillerValasekAdventures, KeenSecurity}.

The vulnerabilities of the in-vehicle networks have already been addressed by cyber-security researchers. One of the first works presenting the vulnerabilities of a licensed vehicle is presented in~\cite{Koscher2010Experimental}, in which the authors experimentally evaluated the security of a modern vehicle. The work presented in~\cite{Koscher2010Experimental} demonstrated the vulnerabilities of poorly-designed systems (such as the Adaptive Cruise Control) and the implications of using legacy communication protocols with no support to modern security techniques. Another similar work is presented in~\cite{MillerValasekBlackHatPaper}, in which the authors focused on the identification of a series of vulnerabilities to remotely hijacking the vehicle. Despite the detailed description of the methodology presented in~\cite{MillerValasekBlackHatPaper} is focused on the steps required to gain remote access to the connected vehicle, the final goal of their work is to inject messages at CAN level, hence demonstrating how it is possible to hijack a connected vehicle via its drive-by-wire capabilities.
In this paper we present first complete powertrain system of a modern vehicle, including both the engine model and CAN communication, that allows to investigate the consequences of cyber-attacks targeting both the internal communication network and the physical representation of the vehicle. Compared to the previous literature, the work presented in this paper presents a generalized approach for the analysis of the consequences of cyber-attacks to a generic engine model, instead of focusing on a particular vehicle.

Modern vehicles are cyber-physical systems (CPSs) \cite{chakraborty2016automotive}. CPSs are systems which show a tight connection between sensing, processing, control and communication to achieve goals that otherwise it would be impossible (e.g. advanced driver assistance system). This scenario offers many attack surfaces where cyber-security tools may not cover \cite{griffioen2019tutorial}.  

One example is the replay attack, where the adversaries replace the actual sensor measurements with pre-recorded ones, while driving the actual physical system to possible dangerous situations. To detect this kind of stealthy attack, system theory tools are needed since the attack is changes the information of the data and not the structure.  

Classical system theory detection technique are based on fault detection schemes which generate a residual obtained from the the actual sensor measurements and the predicted ones by an observer, is given to an anomaly detector. In case of stealthy attacks like the replay attack, this scheme is not able to detect it since 
since the actual measurements are corrupted in a such a way that the anomaly detector cannot detect the attack. For the replay attack an active detection method, watermarking, has been proposed in~\cite{mo2009secure}. The replay attack represents the typology of attack that does not need any knowledge of the dynamical system, as Denial of Service (DoS) attack, and eavesdropping attack~\cite{teixeira2012attack}.

Other interesting works are presented in~\cite{Tianxiang2017DoSCPS, Zhang2021DoSPredictive}, in which a Denial-of-Service attack against a vehicle system is presented. While in~\cite{Tianxiang2017DoSCPS} the authors focus on the analysis of the consequences of the missing input of a generic engine drive, the authors of~\cite{Zhang2021DoSPredictive} designed a predictive controller to overcome missing input. Both~\cite{Tianxiang2017DoSCPS, Zhang2021DoSPredictive} are based on a logical interruption of the input to the systems, which are represented by an approximation of the whole vehicle system instead of considering the actual mechanical parts composing the engine model.
However, compared to both~\cite{Tianxiang2017DoSCPS, Zhang2021DoSPredictive}, we showcase the consequences of a wider range of cyber-attacks, carried out on a model representing the actual mechanical composition of a modern internal combustion engine, and used a simulated communication network to study the realistic consequences of cyber-attacks targeting our system. Moreover, we adapted the replay attack presented in~\cite{mo2009secure} to the automotive scenario to present its consequences on a novel scenario.

\section{Powertrain Model}
\label{s:powertrain}
In this section the basic knowledge required for the understanding of this paper is provided. In Section~\ref{ss:math_mdl} we describe the dynamical model of the IC engine, while in Section~\ref{ss:ss_controller} we present the design of the controller deployed on the IC engine model. Finally, in Section~\ref{ss:can} the basics of the Controller Area Network and on its design in our model are provided.
 
\subsection{Engine and Controller Models}
\label{ss:math_mdl}
We consider the mean-value model (MVM) of the sparkle-ignited (SI) engine, which consists of five interconnected sub-modules: throttle body, intake manifold, gas exchange, combustion and torque generation, and engine inertia. 
The nonlinear model that describes the dynamics of the SI engine is:  
\begin{eqnarray}
\frac{d p_m(t)}{dt}&=&\frac{R\theta_m}{V_d}(\dot m_{\alpha}(t)-\dot m_{\beta}(t)) \label{dpm_dt}\\ 
\frac{d \omega_e(t)}{dt}&=&\frac{1}{\theta_e}[T_e(t)-T_l(t)] \label{dwe_dt}
\end{eqnarray}
where $p_m$ is the intake manifold pressure, $\dot m_{\alpha}$ is the air-flow mass entering from the throttle valve, $\dot m_\beta$ is the air-flow mass going out from the intake manifold to the cylinders, $w_e$ is the engine speed, $T_e$ the torque generated by the engine, and $T_l$ the load torque. 
To use the aforementioned model for the design of a controller, we need to identify the relationship between the two dynamics described in \eqref{dpm_dt} and \eqref{dwe_dt}. Starting from \eqref{dpm_dt} we have to characterize $\dot m_\alpha$ and $\dot m_\beta$. Assuming  that the air is a perfect gas and the throttle is isenthalpic\footnote{the temperature of the air flow in input is approximately the same of the air flow in output}, $\dot m_{\alpha}$ is given by:
\begin{equation}
\dot m_\alpha(t)=\left\lbrace\begin{array}{ll}
A_{\alpha}(t)\frac{p_a}{\sqrt{R\theta_a}}\frac{1}{\sqrt{2}} & \frac{p_m(t)}{p_a}\leq 0.5\\
A_{\alpha}(t)\frac{p_a}{\sqrt{R\theta_a}}\sqrt{\frac{p_m(t)}{p_a}[1-\frac{p_m(t)}{p_a}]} & else
\end{array} \right. \label{dot_ma}
\end{equation}

where $A_{\alpha}$ is the throttle valve open area that is computed as follows:
\begin{equation}
A_{\alpha}(\alpha_{th})=\frac{\pi d_{th}^2}{4}\left(\frac{\cos(\alpha_{th})}{\cos(\alpha_{th,0})}\right)+A_{th,leak} \label{Aa}
\end{equation}

\begin{equation}
\alpha_{th}=\alpha_{th,0}+\left(\frac{\pi}{2}-\alpha_{th,0}\right) u_\alpha \label{alphath}
\end{equation}

where $\alpha_{th}$ is the throttle angle, and $u_{\alpha} \in [0, 1]$ is the control input.

Equations \eqref{Aa} and \eqref{alphath} assume that the throttle actuation is neglected. 
The following relation:
\begin{equation}
\lambda(t)=\frac{1}{\sigma_0} \cdot \frac{\dot m_{\beta}(t)}{\dot m_{\phi}(t)}
\end{equation}
describes the ratio of  air and fuel in the cylinders with respect to the stoichiometric constant $\sigma_0$\footnote{represents the optimal air/fuel ratio which generates the maximum heat possible, i.e. for gasoline is $14.66$.}. 
The term $\dot m_{\phi}$ is the fuel mass flow to the cylinder. Assuming that the injectors dynamics and wall-wetting phenomena are negligible and $\lambda$ is approximately constant, we can derive the expression of the fuel flow-mass going into the cylinder as:
\begin{equation}
\dot m_\phi=\frac{\dot m_{\beta}(t)}{\sigma_0 \lambda} \label{dot_mphi},
\end{equation}
hence the gas-mixture mass flow aspired in the cylinder is:
\[
\dot m(t)=\dot m_{\beta}+ \dot m_{\phi}=\dot m_\beta \left( 1+\frac{1}{\sigma_0 \lambda(t)}\right).
\]
Finally, the air flow-mass going out from the intake manifold to the cylinders $\dot m_\beta$ can be expressed in terms of $\dot m(t)$ as follows:
\begin{equation}
\dot m_{\beta}(t)=\frac{\dot m(t)}{1+\frac{1}{\lambda \sigma_0}} \label{dot_mb}
\end{equation}

Equation~\eqref{dot_mb} presents the relation between $\dot m_\beta$ and $\omega_e$, considering the engine as a volumetric pump, the gas-mixture mass flow aspired in the cylinders is given by:
\begin{eqnarray}
\dot{m}(t)&=&\frac{p_m(t)}{R\theta_m}\cdot \lambda_l(\omega_e(t),p_m(t))\cdot V_d \cdot \frac{\omega_e(t)}{4\pi}
\end{eqnarray} 
where $\lambda_l$ is the volumetric efficiency\footnote{describes how far the engine differs from a perfect volumetric device}. The volumetric efficiency $\lambda_l$ can be approximated as:
\begin{equation}
\lambda_l(\omega_e,p_m)=\lambda_{l\omega}(\omega_e)\cdot \lambda_{lp}(p_m) 
\end{equation}
where
\begin{eqnarray}
\lambda_{l\omega}(\omega_e)&=&\gamma_0+\gamma_1\omega_e+\gamma_2 \omega_e^2\\
\lambda_{lp}(p_m)&=&\frac{V_c+V_d}{V_d}-\frac{V_c}{V_d}\left(\frac{p_{out}}{p_m}\right)^{\frac{1}{\kappa}} \label{lambda_lppm}\end{eqnarray}

In Equation~\eqref{lambda_lppm}, $V_c$ is the compression volume, $V_d$ is the volume displacement, $\kappa$ is the ratio of the specific heat ($\kappa \approx 1.4$), and $p_{out}$ is the pressure at the engine's exhaust side. 
Now we know how $\omega_e$ is related to \eqref{dpm_dt}, and to show how $p_m$ is related to \eqref{dwe_dt} we introduce the break mean effective pressure $p_{me}$ which represents the required pressure on the piston during one full expansion stroke to complete an engine cycle\footnote{for a four-stroke engine it corresponds to two engine revolutions.} and it is given by:
\begin{equation}
p_{me}=\frac{T_e \cdot 4\pi}{V_d}. \label{pme}
\end{equation}

The fuel mean effective pressure:
\begin{equation}
p_{m_\phi}=\frac{m_\phi \cdot H_l}{V_d},
\end{equation}
corresponds to break mean effective pressure generated by an engine that converts all the fuel thermal energy into mechanical energy, for the amount of fuel mass $m_\phi$ burnt per engine cycle. $H_l$ is the specific energy of the fuel. The mass of fuel $m_\phi$ is related to the fuel mass flow by the following equation: 
\begin{equation}
\dot m_{\phi}(t)=m_\phi(t)\frac{w_e(t)}{4\pi}. \label{dot_mphi1}
\end{equation}
This equation also represents the relationship that links the mechanical dynamics with the air pressure in the intake manifold. 

To exploit this relationship, we start considering the effective efficiency which is:
\begin{equation}
\eta_e=\frac{p_{me}}{p_{m_\phi}}=\frac{T_e \cdot 4\pi}{m_\phi \cdot H_f} \label{pmphi}
\end{equation}
and the goal is to find an expression for $T_e$. Hence then:
\begin{equation}
p_{me}=\eta_e(....) \cdot p_{m_\phi} \label{pme1}
\end{equation}
$\eta_e(...)$ can be obtained in different ways, a possible approximation uses the indicated mean pressure:
\begin{equation}
p_{me}\approx p_{mi}(\omega_e)-(p_{m0f}(\omega_e)+p_{m0g}(\omega_e)) \label{pme2}
\end{equation}
where $p_{mi}$ is the indicated mean pressure which is given by:
\begin{equation}
p_{mi}=\frac{w_i}{V_d}=\eta_i \frac{m_\phi \cdot H_f}{V_d}
\end{equation}
where $w_i$ is the indicated work\footnote{mechanical energy transferred to the piston during one cycle, where friction is not considered.}. The associated indicated thermodynamic efficiency can be approximated by using the \emph{Willians} approximation as: $\eta_i(\omega_e)\approx \eta_0 + \eta_1 \omega_e$. The terms $p_{m0f}$ and $p_{m0g}$ represent the loss due the friction and gas exchange.

From \eqref{pme}, \eqref{pme1}, and \eqref{pme2}
\begin{eqnarray}
T_e(t)&=&p_{me}(t)\cdot \frac{V_d}{4\pi}\nonumber\\ 
&=&(\eta_i(\omega_e)p_{m_\phi}-p_{m0f}(\omega_e)-p_{m0g}(\omega_e))\frac{V_d}{4\pi}
\end{eqnarray}

From \eqref{pmphi} we find the expression of $p_{m_\phi}$, hence:
\begin{equation}
T_e(t)=((\eta_0 + \eta_1 \omega_e)\frac{H_f\cdot m_\phi(t)}{V_d}-p_{m0f}(\omega_e)-p_{m0g}(\omega_e))\frac{V_d}{4\pi} \label{Te}
\end{equation}

By using \eqref{dot_mphi1}, we can express $m_{\phi}$ as function of $\dot m_{\phi}$ which can be derived from \eqref{dot_mphi} as:
\begin{equation}
m_{\phi}(t)=\frac{\dot m_{\beta} \cdot 4\pi}{\alpha \cdot \omega_e}\footnote{for idle control gas-mixing transportation delays have to be taken into account, $\dot m_\phi(t-\delta)$ and $\dot m_\beta(t-\delta)$.} \label{mphi}
\end{equation}
where $\alpha=\lambda \cdot \sigma_0$. 

\begin{figure*}
\begin{center}
\begin{eqnarray}
\hspace*{-0.3cm}\frac{dp_m(t)}{dt}&=&\frac{R\theta_m}{V_d}\left(A_\alpha(t)\frac{p_a}{\sqrt{R\theta_a}}\frac{1}{\sqrt(2)}-\left(\frac{p_m(t)}{R\theta_m}\left(\gamma_0+\gamma_1 \omega_e(t)+\gamma_2 \omega_e^2(t)\right)\right.\right. \nonumber \\
&&\left. \left.\left(\frac{V_c+V_d}{V_d}-\frac{V_c}{V_d}\left(\frac{p_{out}}{p_m}\right)^{\frac{1}{\kappa}}\right)\frac{V_d\omega_e(t)}{4\pi}\frac{\alpha}{\alpha+1} \right) \right) \label{dpm1} \\
\frac{d\omega_e(t)}{dt}&=&\frac{1}{\theta_e}\left[\left(\left(\eta_0+\eta_1\omega_e(t) \right)\frac{H_f \cdot p_m(t)}{R\theta_m}\left(\gamma_0+\gamma_1 \omega_e(t)+\gamma_2 \omega_e^2(t)\right)\left(\frac{V_c+V_d}{V_d}-\frac{V_c}{V_d}\left(\frac{p_{out}}{p_m}\right)^{\frac{1}{\kappa}}\right)\right.\right. \nonumber\\
&&\left.\left. \cdot \frac{V_d}{\alpha+1}-\left( \beta_0+\beta_2\omega_e^2(t)+(p_{out}-p_m(t)) \right)\frac{V_d}{4\pi} \right)-T_l(t)  \right] \label{dwe1}
\end{eqnarray}
\caption{Nonlinear model of an SI-engine for cruise-control problem.}
\label{fig:nonlinear_model}
\end{center}
\end{figure*}

\begin{table}[htpb]
\centering
\begin{tabular}{@{}lrr|lrr@{}}
\toprule
\textbf{param} & \textbf{value} & \textbf{units} & \textbf{param} & \textbf{value} & \textbf{units} \\ 
\midrule
$R$ & 287 & [J/KgK] & $\gamma1$ & 3.42e-3 & [s] \\
$\theta_a$ & 298 & [K] & $\gamma2$ & -7.7e-6 & [$s^2$] \\
$\theta_m$ & 340 & [K] & $\eta0$ & 0.16 & [J/Kg] \\
$\alpha_{th0}$ & 7.9 & [deg] & $\eta1$ & 2.21e-3 & [Js/Kg] \\
$d_{th}$ & 58.7e-3 & [m] & $\beta0$ & 15.6 & [Nm] \\
$A_{th,leak}$ & 5.6e-6 & [$m^2$] & $\beta2$ & 0.175e-3 & [$Nms^2$] \\
$Vd$ & 2.77e-3 & [$m^3$] & $\theta_e$ & 0.2 & [kg/$m^2$] \\
$Vc$ & 0.277e-3 & [$m^3$] & $H_f$ & 45.8e6 & [-] \\
$p_a$ & 1e5 & [Pa] & $\kappa$ & 1.35 & [-] \\
$p_{out}$ & 1e5 & [Pa] & $\alpha$ & 14.70 & [-] \\
$\gamma0$ & 0.45  & [-] & & & \\
\bottomrule
\end{tabular}
\caption{Values of the parameters used in the model}
\label{tab:model_param}
\end{table}

Fig. \ref{fig:nonlinear_model} shows the nonlinear model we consider to simulate and control the SI IC engine, where equation \eqref{dpm_dt} has been rewritten by using \eqref{dot_ma} and \eqref{dot_mb}, and equation \eqref{dwe_dt} with \eqref{Te} and \eqref{mphi}. Table~\ref{tab:model_param} shows the parameters we use in our simulations \cite{guzzella2009introduction}.

\subsection{State Space Representation and Controller Design}
\label{ss:ss_controller}
In this section we present the controller design for our SI engine described by \eqref{dpm1} and \eqref{dwe1}. In this model we can identify $p_m$ and $w_e$ as state variables:
\begin{equation}
z \triangleq \left[ \begin{array}{c}
p_m\\
w_e
\end{array}\right],
\end{equation}

where $h(t)=A_{\alpha}(t)$\footnote{To do not overload the representation we consider directly $A_{\alpha}(t)$ as input. To consider $u_\alpha$ the throttle angle, we need to consider the equations \eqref{Aa} and \eqref{alphath}.} as the input, and $w_e$ is the output. 
The cruise control problem aims to keep a certain constant speed of the car which means that the SI engine is working around a pre-determined equilibrium point $\bar z = [\bar p_m, \bar w_e]^T$ and $\bar{h}=\bar A_\alpha$.

A controller for this task can be designed considering the linearized dynamics \eqref{dpm1} and \eqref{dwe1} around $\bar{x}$. Considering a constant $T_l$\footnote{for a case of replay attack, a time-varying $T_l$ makes the attack not stealthy, here we consider a scenario where $T_l$ is constant and possible small changes can be modelled as random noise.}, \eqref{dpm1} and \eqref{dwe1} can be written as $\dot{z}=f(z,h)$, where $f:\mathbb{R}^2\times \mathbb{R}\rightarrow \mathbb{R}^2 $. From the linearization:
\begin{equation}
    A= \frac{\partial f}{\partial z}\bigg\rvert_{z=\bar{z}\; h=\bar{h}},\quad B= \frac{\partial f}{\partial h}\bigg\rvert_{z=\bar{z}\; h=\bar{h}}  
\end{equation}
and defining $x \triangleq z-\bar{z}$ and $u \triangleq h-\bar{h}$, the system we consider for design the controller is:
\begin{equation}
    \dot x = Ax+Bu+w \label{lin_model_ct}
\end{equation}
where $w \sim \mathcal{N}(0,\,Q)$ represents the process noise which is zero-mean Gaussian with known covariance matrix $Q$. Since the output is $w_e$, then the output equation of the linearized model is $y=Cx+v$ with $C=[0\; 1]$ and $v \sim \mathcal{N}(0,\,R)$ is the zero-mean Gaussian measurement noise with variance $R$. To design the controller we consider the discrete version of \eqref{lin_model_ct} with sampling time $T_s$. Hence the considered system is given by:
\begin{eqnarray}
x_{k+1}&=& A_dx_k+B_d u_k+w_k \label{lin_dynamics_dt}\\
y_k &=&Cx_k+v_k \label{lin_output_dt}
\end{eqnarray}
with $w_k \sim \mathcal{N}(0,\,Q_d)$, and $v_k \sim \mathcal{N}(0,\,R_d)$. We defined $A_d$, $B_d$, $Q_d$, and $R_d$ as in~\cite{yaghooti2021physical}. 
To stabilize the dynamics~\eqref{lin_dynamics_dt}, we consider a Linear Quadratic Gaussian (LQG) controller. This is a state feedback controller $u_k=Lx_k$ that minimizes the following cost function:
\begin{equation}
    J = \lim_{N\rightarrow\infty}\mathbb{E}\frac{1}{N}\bigg[\sum_{k=0}^{N-1}(x_k^\top Wx_k + u_k^\top Uu_k) \bigg],
\end{equation}
where $W$ and $U$ are semi-positive definite matrices. The actual control is computed as:
\begin{equation}
    u_k=L\hat x_{k|k} \label{LQR_control}
\end{equation}
where $\hat x_{k|k}$ is the Kalman state estimation at time $k$
\begin{equation}
    \hat x_{k|k} = \hat{x}_{k|k-1}+K\left(y_k-C\hat{x}_{k|k-1} \right). \label{Kalman_filter}
\end{equation}

The one-step prediction $\hat{x}_{k|k-1}$ is given by:
\begin{equation}
    \hat{x}_{k|k-1}=A_d\hat{x}_{k-1|k-1}+B_d u_{k-1}.
\end{equation}
The constant gain $K$ is the Kalman gain considered at steady state. For more details about \eqref{LQR_control} and \eqref{Kalman_filter}, the reader is invite to see~\cite{yaghooti2021physical}. The associated observer-based state feedback control scheme is reported in Fig. \ref{fig:sys_model}.

\subsection{The Controller Area Network}
\label{ss:can}
\begin{figure}[htb]
    \centering
    \includegraphics[width=0.48\textwidth]{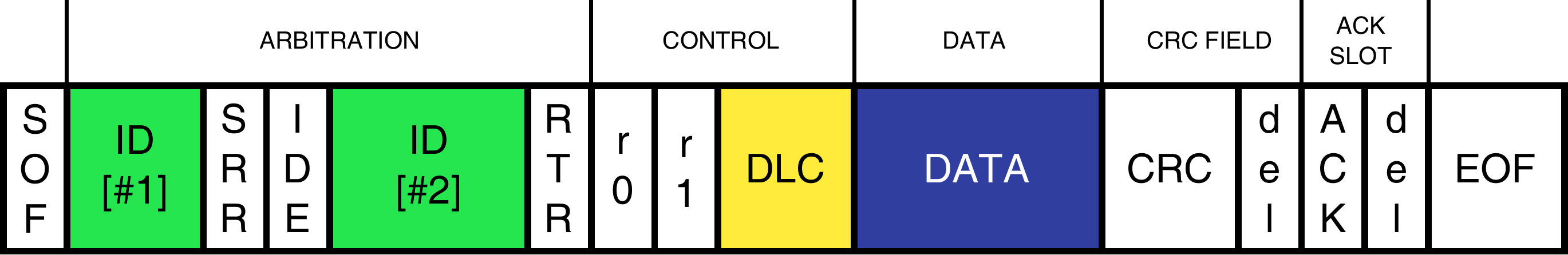}
    \caption{CAN data frame in the extended format}
    \label{fig:data_frame_extended}
\end{figure}

The Controller Area Network (CAN) is a vehicle bus standard designed to allow the nodes of the network to exchange data without requiring a host computer~\cite{BOSCHCanSpecV2}. CAN is one of the most deployed networking protocol for internal vehicular communications due to its high resilience to electromagnetic interference and its cheap implementation. 
Microcontrollers on the same CAN segment exchange data between themselves using the CAN data frame, one of the $4$ types of frames defined by the CAN standard. 
The CAN data frame is composed by two main fields, namely the \emph{identifier (ID)}, and the \emph{payload (data)}. The \emph{ID} is used to distinguish among different types of CAN data frame. Data frames characterized by a given ID are produced by only one microcontroller, while receiver microcontrollers use the value of the ID to select data frames that are relevant for their functioning. The CAN standard defines two types of data frames: the \emph{standard} format, whose ID field is $11$-bits long, and the \emph{extended} format, whose ID field is encoded using $29$-bits. The extra bits of the extended format are separated from the bits composing the standard format for backward compatibility. Figure~\ref{fig:data_frame_extended} shows an example of a generic CAN data frame in the \emph{extended} format.
The \emph{data} field encapsulates the information that the sender microcontroller transmits to other microcontrollers on the network. The data field has a variable size (from $1$ to $8$ bytes) and usually packs several different signals. The CAN standard leaves complete freedom to the car manufacturers about the structure, number, encoding, and semantic of these signals. 
Hence, without having access to the formal specifications of the CAN network for a particular vehicle model, the signals encoded in the data field can only be interpreted as an opaque binary blob.

For the tests presented in this paper, we designed a CAN communication network between the IC engine model the speed controller. In particular, we identified a minimum of two messages required to enable communication between the IC engine model and the speed controller. The first message, called \textit{ES (engine speed)} is sent from the ECU attached to the \emph{engine output sensor} and required as input by the \emph{controller}, while the second message is called \textit{TR (throttle request)} and is sent from the \emph{controller} to the input of the \emph{engine model}. The \textit{ES} message contains the value of the engine rotational speed, while the \textit{TR} message contains the value of the throttle estimated by the controller to keep the rotational speed constant. 
The detailed description of the two CAN messages is provided in Table~\ref{tab:can_msgs}.

\begin{table}[htbp]
\centering
\begin{tabular}{@{}ccccc@{}}
\toprule
\textbf{Message} & \textbf{ID {[}hex{]}} & \textbf{cycle time {[}ms{]}} & \textbf{encoded signal} & \textbf{DLC {[}bits{]}} \\ \midrule
ES & 0x10 & 10 & \begin{tabular}[c]{@{}c@{}}throttle opening\\ request {[}0, 1{]}\end{tabular} & 32 \\
TR & 0x15 & 10 & \begin{tabular}[c]{@{}c@{}}engine rotational\\ speed {[}RPM{]}\end{tabular} & 32 \\ 
\bottomrule
\end{tabular}\\
\caption{CAN bus specifications}
\label{tab:can_msgs}
\end{table}

\begin{figure}[hptb]
    \centering
    \includegraphics[width=.95\columnwidth]{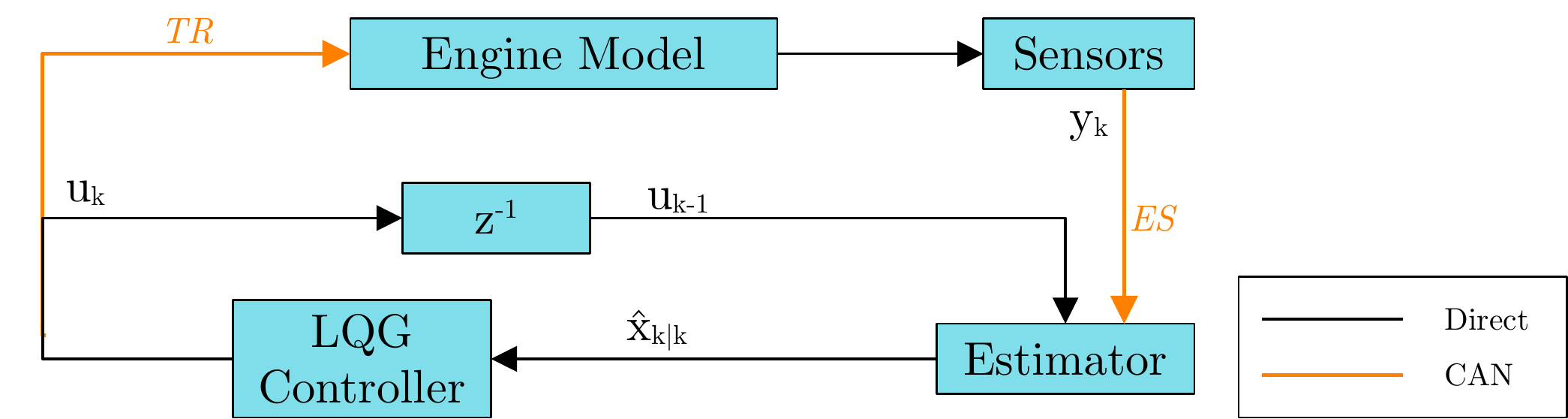}
    \caption{Final design of the IC engine model with the controller and CAN communication}
    \label{fig:sys_model}
\end{figure}

Figure~\ref{fig:sys_model} presents the final architecture of the system, showing how the different subsystems are connected with each other: a black line is used to denote direct connection between the subsystems, while the orange line is used to highlight subsystems communicating via CAN.

\section{Threat Model}
\label{s:threat_model}
In this section we present the threat model considered for testing our model. The threat model considered in this paper is focused on attacks targeting the output of the engine model, with the final goal to modify its output.


Modern literature already showcased different threat models for both CPS security and CAN communications, demonstrating the vulnerabilities of modern control systems and networking protocols. The consequences of these vulnerabilities have been exploited by security researchers from both academia and industries to perform different attacks, from simple \textit{Denial-of-Services}~\cite{Tianxiang2017DoSCPS, Zhang2021DoSPredictive}, designed to test the systems against a disruption of the communication, to more advanced \textit{impersonation attacks}~\cite{nowdehi2019casad}, in which the attacker is able to impersonate a target ECUs by replacing legit messages with maliciously forged ones.
However, all the presented threat models are developed by focusing only on a specific aspect of the model, being either the control system itself~\cite{Schmittner2015FMVEA, Jamil2021Towards} or the adopted communication protocol~\cite{checkoway2011comprehensive, McCune2016Simulation}. 

The threat model considered in this paper is based on both research areas, hence representing a hybrid approach composed by control system theory and CAN communication. To the best of our knowledge, this is the first time a hybrid approach is used for the definition of a threat model for automotive applications.
The threat model considered in this paper is composed by three different attack scenarios, each one designed to modify the output of the system by means of different attack vectors. The detailed description of each attack scenario is presented in the following.

\subsection{Fuzzing attack}
\label{s:threat_fuzzing}
The fuzzing attack considered in our threat model requires an attacker with the ability to provide custom input to the control system, resulting in the modification of the speed of the engine.
This attack supposes that the attacker is able to inject a custom control input to the system at anytime. The attack is composed by two stages: in the first stage the attacker learns the legit values of the input of the control system, while in the second stage a desired sequence of control input is provided to the system. 
The fuzzing attack is a type of attack already explored on both control systems~\cite{Chen2019Fuzzing} and CAN communications~\cite{Lee2015Fuzzing, Martinelli2017Fuzzing}. However, since the input of the system can be modified at sensor level or by tampering the content of its relative CAN message, we remark that the consequences of the fuzzing attack scenario can be analyzed by targeting either the model's sensors or the CAN communication.

\subsection{Replay attack}
\label{s:threat_replay}
The replay attack is an advanced version of the \emph{fuzzing attack} scenario, in which it is necessary that the attacker is able to provide custom input to the control system while simultaneously change its output by replaying the corresponding output of the provided input.
This attack scenario is composed by two stages: in the first stage the attacker records a sufficient number of sensor readings (both input and output) without providing any input to the system, hence with the system running in normal conditions; while in the second stage the attacker provides a desired sequence of control input while replaying the previously recorded outputs. By conducting a replay attack to the powertrain system, the attacker is able to modify the final speed of the vehicle without being noticed, by injecting a control input to the system and masquerading its consequences to the controller by replaying previously recorded output values. This attack scenario has already explored on both research areas~\cite{Mo2009Replay, Groza2019Efficient} hence we remark that its consequences on our model can be analyzed by simulating the attack on one of the two scenarios.


\subsection{Injection attack}
\label{s:threat_injection}
In the injection attack scenario, the attacker has no direct access to the sensors attached to the ECUs but is able to read and send CAN messages, either via a physical connection or by exploiting a vulnerability of a target ECU. The injection attack is an attack targeting the CAN communication between the system microcontrollers, in which the attacker is able to send messages containing an arbitrary value of the system control input to modify its output. The injection attack is designed in two stages: in the first stage the attacker observes the CAN communication of the target vehicle to identify the messages carrying the input values of the control system and to learn their cycle time; while in the second stage the attacker injects CAN messages with the desired control input encoded in their data field with a faster cycle time than the original ones, interleaving malicious messages with the valid ones. As opposed to the previous two attack scenarios, the injection attack can only be performed at CAN communication level~\cite{nowdehi2019casad}.

\section{Attack consequences}
\label{s:attack_consequences}
In this section we investigate the consequences of the attacks composing the threat model presented in Section~\ref{s:threat_model} on the powertrain model presented in Section~\ref{s:powertrain}.
The attacks are replicated on our model after $10$ seconds of normal system simulation for a duration of $2$ seconds.

\subsection{Fuzzing attack}
\label{ss:consequence_fuzzing}
\begin{figure}[hpbt]
    \centering
    \includegraphics[width=.8\columnwidth]{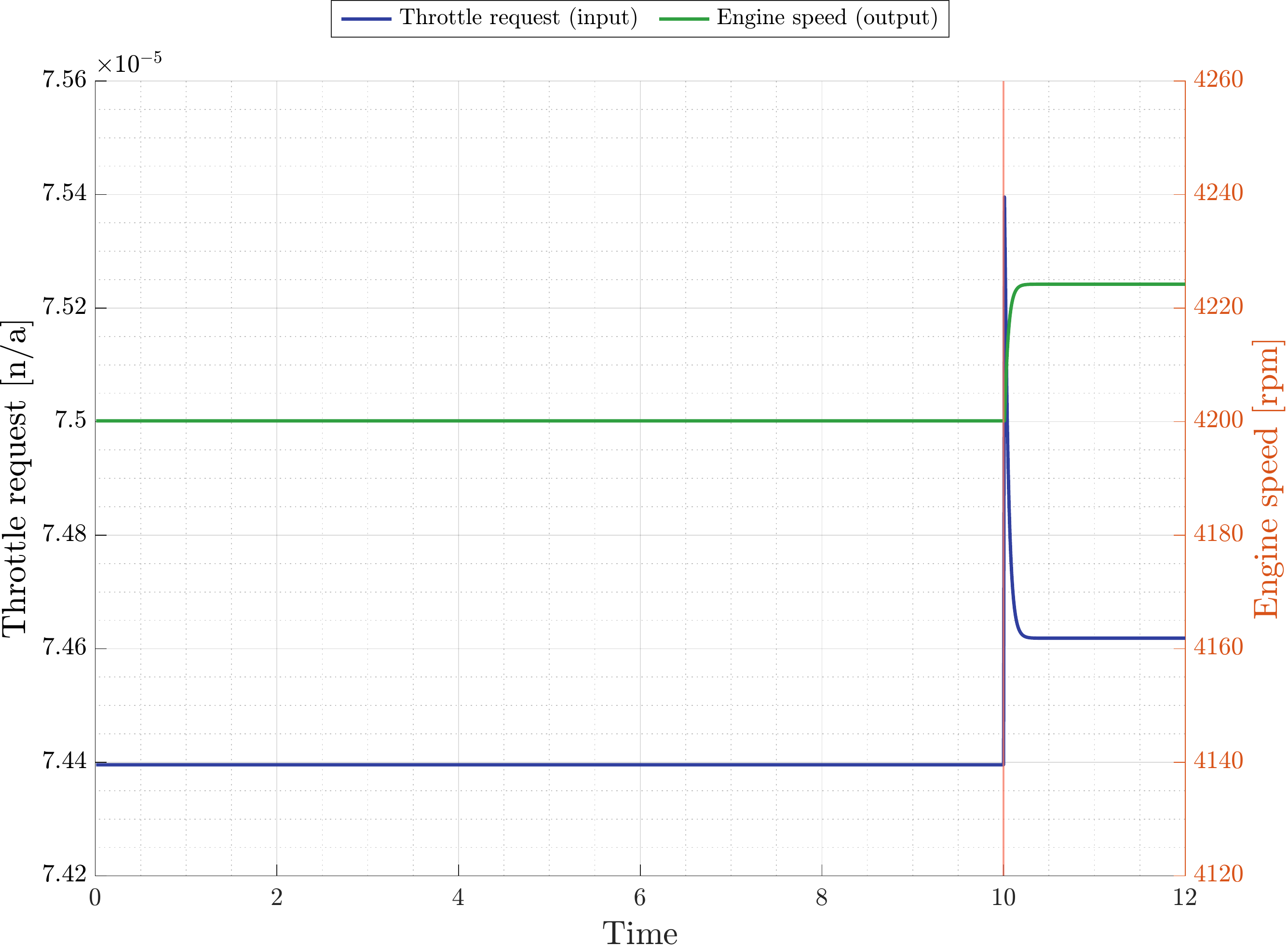}
    \caption{Consequences of the fuzzing attack on the system input}
    \label{fig:fuzzing_consequences}
\end{figure}

The consequences of the fuzzing attack on our model are investigated by performing the attack at CAN level. 
The attack is simulated as follows. In the first phase of the attack we identified a sequence of input values corresponding to $2$ seconds of system inputs. In the second phase we changed the real value of the sensor reading encoded in its relative CAN message by increasing the values of a fixed amount. The attack analyzed in this section targets the input of the engine model (the throttle request), and the value encoded in the CAN messages is increased by $1e-6$.
Figure~\ref{fig:fuzzing_consequences} shows the consequences of this attack on our system, comparing the input (throttle request, blue line) with the output (engine speed, green line). The left $y$-axis shows the values of the input, the right $y$-axis shows the values of the output, while the $x$-axis shows the time of the simulated system. We denote the start of the attack with a red vertical line.

From the analysis of the results presented in Figure~\ref{fig:fuzzing_consequences} we notice that after the start of the attack the output of the system rapidly increased, from its stationary value of $4200$rpm to a peak of $4224$rpm. Moreover, despite the input of the system is fixed at $7.54e-5$, the action of the cruise controller deployed on our model stabilizes the input of the system to approximately $7.462e-5$, hence slightly decreasing the malicious input. 

\subsection{Replay attack}
\label{ss:consequence_replay}
\begin{figure}[hpbt]
    \centering
    \includegraphics[width=.8\columnwidth]{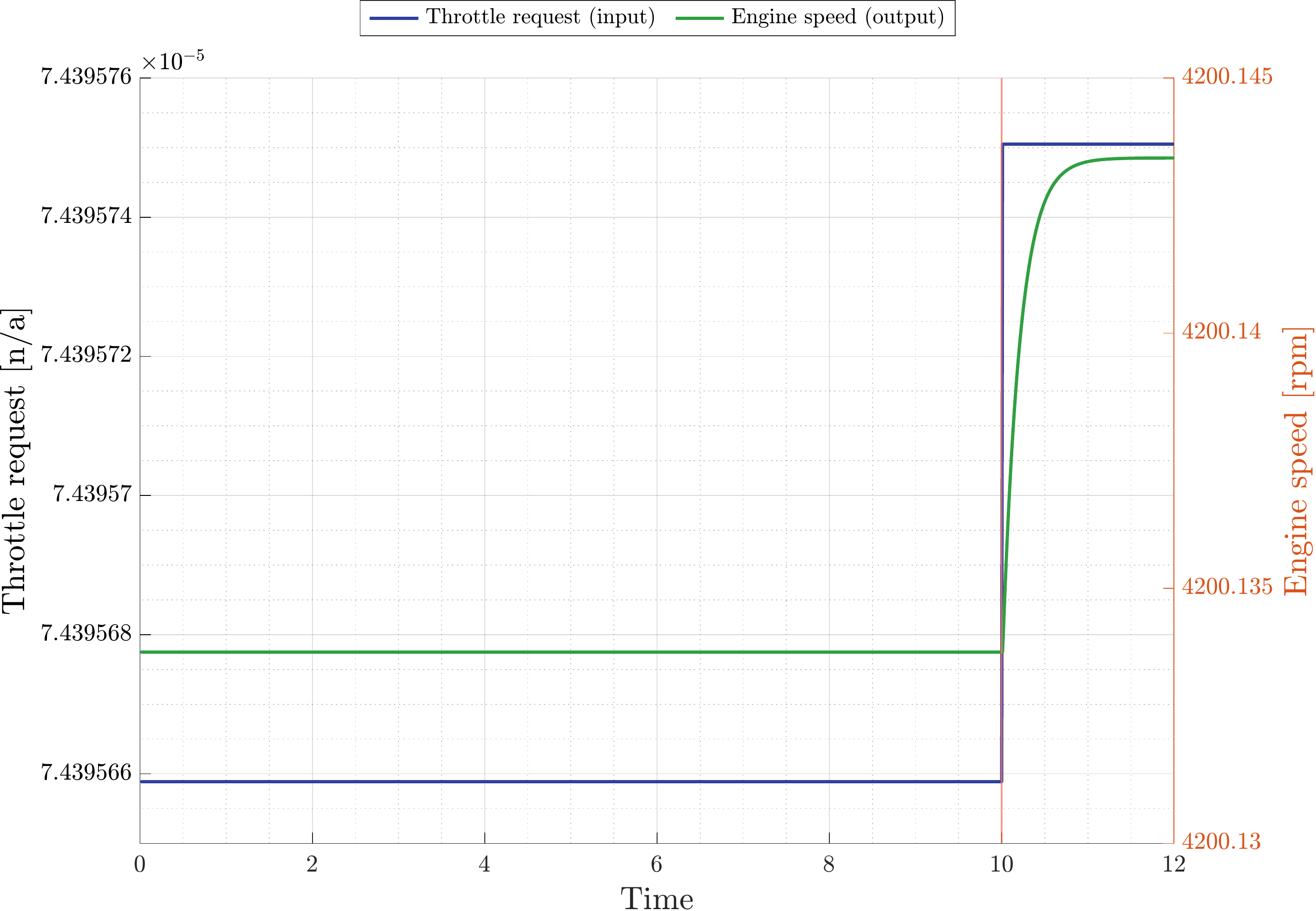}
    \caption{Consequences of the replay attack on the system}
    \label{fig:replay_consequences}
\end{figure}

The consequences of the replay attack on our model are investigated by performing the attack at CAN level. For this attack scenario, we also remark that the same results are achieved by performing the same attack at sensor level.  
The attack is conducted as follows. In the first phase of the attack we recorded a sequence of input and output values corresponding to $2$ seconds. The second phase of the attack is conducted as in the \emph{fuzzing attack} scenario, despite in this attack scenario the content of the CAN message carrying the reading's of the output sensor is overwritten with the output value corresponding to the given input.
Figure~\ref{fig:replay_consequences} shows the consequences of this attack on our system, comparing the input (throttle request, blue line) with the output (engine speed, green line). The left $y$-axis shows the values of the input, the right $y$-axis shows the real output of the system (and not the replayed values), while the $x$-axis shows the time of the simulated system. The start of the attack is represented by a red vertical line. 

Figure~\ref{fig:replay_consequences} shows that after the start of the replay attack there is an increment of the actual speed of the engine. Moreover, by comparing the evolution of the input of the system with the results presented in Figure~\ref{fig:fuzzing_consequences}, it is possible to notice that by overwriting the output of the system before its relative CAN message is sent to the controller it is possible to prevent the controller to operate on the input.

\subsection{Injection attack}
\label{ss:consequence_injection}
The consequences of the injection attack scenario are investigated by performing the attack at CAN level. We remark that this attack scenario is limited to the CAN communication bus, hence it is not possible to obtain the same results by performing the attack at sensor level. The injection attack is performed by injecting throttle request CAN message with a higher value of the throttle input to increase the speed of the engine.
The attack is conducted as follows. In the first phase the legit values of the input are analyzed to identify the normal value used by the controller. In the second phase, messages carrying a higher value of the input of the system are injected with a frequency $10$ times higher than the legit message, thus injecting $10$ malicious messages between two legit ones. The input value of the injected message is $1e-6$ higher than the value sent by the original messages.

\begin{figure}[hpbt]
    \centering
    \includegraphics[width=.8\columnwidth]{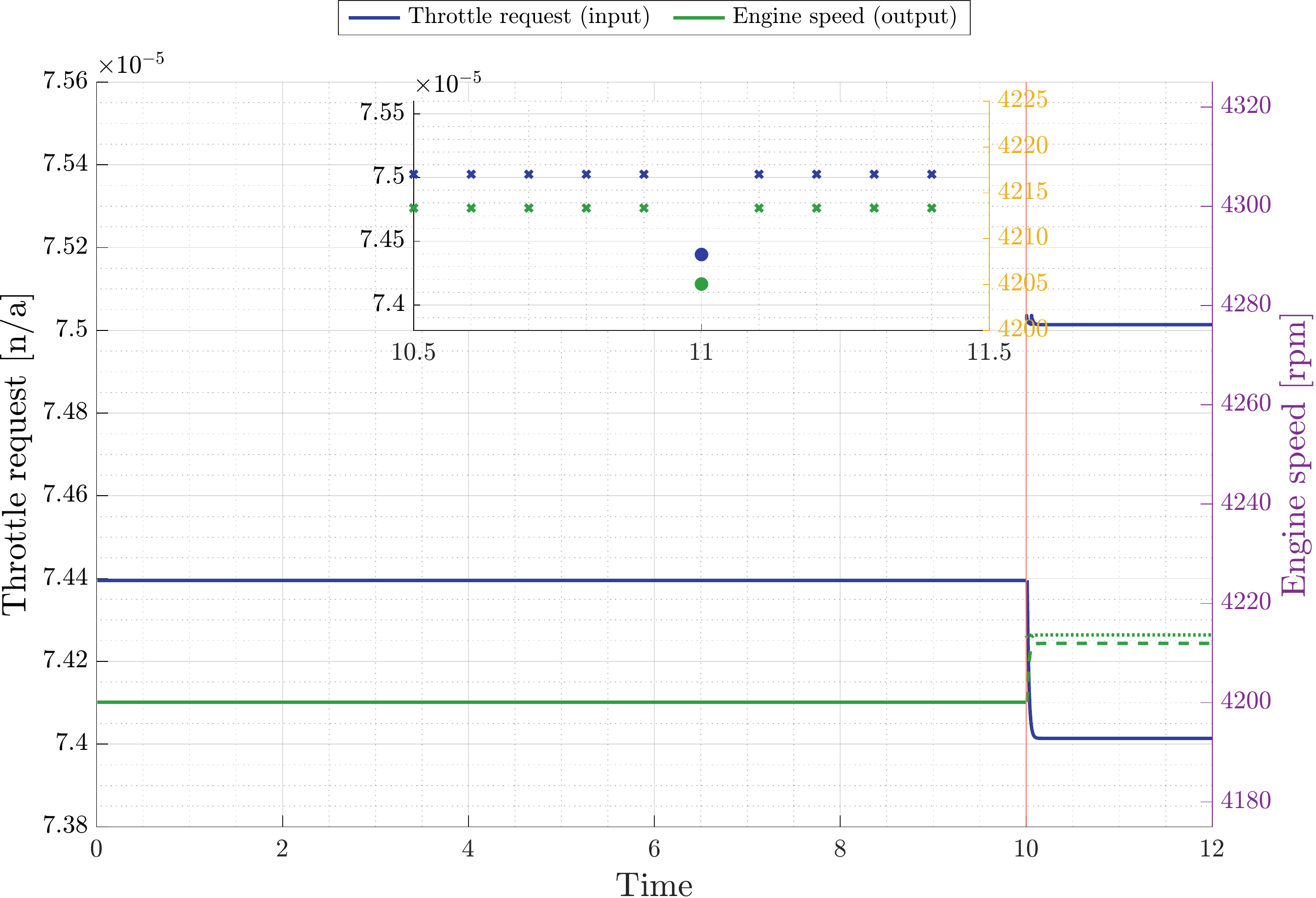}
    \caption{Consequences of the injection attack on the system}
    \label{fig:injection_consequences}
\end{figure}

Figure~\ref{fig:injection_consequences} shows the consequences of this attack on our system, comparing the input (throttle request, blue line) with the output (engine speed, green line). The left $y$-axis shows the values of the input, the right $y$-axis shows the output of the system, while the $x$-axis shows the time of the simulated system. The start of the attack is represented by a red vertical line. The internal plot of Figure~\ref{fig:injection_consequences} is a detailed example of the system input and output centered at time $t=11$ seconds. In this internal Figure it is possible to notice better the difference of the between the legit values of the input (blue dot) and output (green dot) sensors, the injected system input (blue crosses) and its relative output (green crosses).

The analysis of the results presented in Figure~\ref{fig:injection_consequences} shows an interesting pattern caused by the start of the attack. In particular, the output of the system goes from a minimum values of $4120$rpm (which to the legit value) to values of $4215$rpm (which corresponds to the injected input). 
\section{Conclusions}
\label{s:conclusions}
In this paper we analyzed the consequences of cyber-attacks to the powertrain model of a generic vehicle. The powertrain model is built by considering a generic internal combustion engine model, composed by $5$ different sub-models, a speed controller, and a communication network based on the CAN protocol connecting the engine model with the controller.
We consider a threat model composed by different attack scenarios, whose final goal is to modify the output of the engine model, representing the rotational speed of the engine. We discussed $3$ different attacks to the system, of which 2 can be deployed at both sensor and CAN level, while the other attack can be only executed at CAN communication level.
We experimentally analyzed the consequences of these attacks on our model through simulation of the different attacks, demonstrating that it is possible to modify the output of the model by introducing little deviations in the content of the CAN messages or in the sensor's readings.
Compared to the current state-of-the-art, the work presented in this paper uses a cruise controller developed to operate in conjunction with the model describing the actual components of a generic internal combustion engine, demonstrates the consequences of cyber-attacks targeting the powertrain system, and uses the CAN communication protocol to enable communication between the engine and the controller.

Future work is focused on the analysis of state-of-the-art defences against the considered threat model to build a framework that exploits the two scenarios to overcome each other limitations.


\small{
\bibliographystyle{IEEEtran}
\bibliography{bibliography}{}
}

\end{document}